\documentstyle[preprint,eqsecnum,prb,aps,epsf]{revtex}
\newfont{\NF}{cmbx10 scaled\magstephalf}

\begin{document}    

%\draft

  \title{\NF~~\\~\\ 
  CURRENT DISTRIBUTIONS AND\\ THE DE HAAS--VAN ALPHEN OSCILLATION\\
  IN A PLANAR SYSTEM OF HALL ELECTRONS}
  \author{K. SHIZUYA}
  \address{Yukawa Institute for Theoretical Physics\\
  Kyoto University,~Kyoto 606-01,~Japan}

 \maketitle
 
\begin{abstract} 
The current distribution is studied for a
finite-width two-dimensional system of Hall electrons, with 
a clear distinction drawn between the equilibrium edge current and 
the Hall current.
It is pointed out that both the distribution and direction of the
equilibrium edge current change dramatically as the number of 
electron edge states increases, and 
that this alternating edge current is
another manifestation of the de Haas--van Alphen effect.
The Hall-current distribution is substantially different from 
the edge current distribution, and it is shown numerically that 
the fast-traveling electrons along the sample edge are not 
the main carriers of the Hall current.
\end{abstract}

\newpage

\setcounter{section}{1}
\noindent
{\bf 1.\ \  Introduction}
\medskip

One basic question inherent to the quantum Hall effect~\cite{V} (QHE) 
is where in the sample the current flows. The traditional 
explanations~\cite{AA,Pr,L,H} for the QHE are based on a picture 
that the Hall current is carried by electrons in the sample bulk 
and regard local disorder as crucial for the formation of visible
conductance plateaus. 
On the other hand, considerable attention has recently been directed 
to another picture,~\cite{MS,SKM,B} the edge-state picture of the QHE, 
where the current is taken to flow in channels along the edges of a
sample.
Experiments~\cite{WFSK} appear to favor interpretation of observed
results in terms of the edge channels but little is known about the
current distribution in real samples so far, except that  
some information is obtained from observed potential 
distributions.~\cite{FS,FKHBW}   
Theoretically the current distribution as well as the potential
distribution has been studied in some
models,~\cite{MRB,HT,OO,CSG,T,KS,AHK,GV} and  
possible connections between the bulk-state and edge-state pictures 
of the QHE have been discussed.~\cite{T,KS}

The purpose of the present paper is to study the current-carrying 
properties of the electrons in the sample bulk and 
near the edges in detail.
In equilibrium a prominent current flow arises along the edges 
of a Hall sample.
It is pointed out that both the distribution and direction of the edge
current change dramatically 
as electron edge states increase in number
and that this alternating edge current is another manifestation of 
the de Haas -- van Alphen (dHvA) effect.
The Hall-current distribution turns out substantially different from 
the equilibrium current distribution, and it is shown numerically 
that the fast-traveling electrons along the sample edges, 
though carrying a large amount of current per electron, are not 
the main carriers of the Hall current.

In Sec.~2 we study the equilibrium current distribution in a 
disorder-free Hall sample of finite width.  
In Sec.~3 we examine the Hall current distribution and 
discuss the transport properties of the edge current.
Sec.~4 is devoted to concluding remarks where 
the effect of disorder is discussed.\\

\newpage

\setcounter{section}{2}
\noindent
{\bf 2.\ \  Current distributions in equilibrium}
\medskip
 
Consider electrons confined to an infinitely long strip of width
$W$ (or formally, a strip bent into a loop of circumference
$L_{x} \gg  W$) in the presence of a uniform magnetic field $B$
normal to the plane, described by the Hamiltonian:
\begin{equation}
H_{0} = {1\over 2}\omega \Bigl\{ \ell^{2}p_{y}^{2}
+(1/\ell^{2})(y - y_{0})^{2}\Bigr\},
\label{hzero}
\end{equation}
written in terms of $\omega \equiv eB/m$, the magnetic length
${\ell}\equiv 1/\sqrt{eB}$ and $y_{0} \equiv p_{x}/(eB)$; 
the Landau-gauge vector potential $(-By,0)$ has been used to
supply a uniform magnetic field. We take explicit account of 
the two edges $y=\pm W/2$, where the wave function is bound 
to vanish.

The eigenstates of $H_{0}$ in the sample bulk are Landau levels with
discrete spectrum $\omega (n + {1\over2})$ labeled by integers
$n=0,1,2,\cdots$, and degenerate in $y_{0}=\ell^{2} p_{x}$;
the $y_{0}$ labels the center-of-mass position of 
each orbiting electron.
The eigenfunctions in the presence of sharp edges are still 
labeled by $N=(n,\, y_{0})$: 
\begin{equation}
\psi_{N}(x,y) = (L_{x})^{-1/2}\,
e^{ixp_{x}}\phi_{N}(y), \label{psiNzero}
\end{equation}
with $\phi_{N}(y)$ given~\cite{MS} by the parabolic cylinder 
functions~\cite{WW} $D_{\nu}(\pm \sqrt{2}(y-y_{0})/\ell)$ for 
electrons residing near the edges $y = \mp W/2$.
The energy spectrum of each Landau level $n$ 
\begin{equation}
\epsilon_{n}(y_{0}) = \omega \left\{\nu_{n}(y_{0}) 
+ {1\over 2} \right\}
\end{equation}
is fixed from the boundary condition $\phi_{N}=0$ at $y=\pm W/2$.
The spectra $\nu_{n}(y_{0})$ are determined numerically
and are functions of a dimensionless combination
$y_{0}^{W}/\ell$:
\begin{equation}
\nu_{n}(y_{0})={{\nu}_{n}}[y^{W}_{0}/\ell],    \label{nuN} 
\end{equation}
where $y_{0}^{W}\equiv  y_{0} + W/2$ refers to the value of $y_{0}$ 
measured relative to $y_{0} = - W/2$ near the ``lower'' edge; 
$y_{0}^{W}\equiv  y_{0} - W/2$ near the ``upper'' edge $y_{0} \sim W/2$.
As shown in Fig.~1, $\nu_{n}(y_{0}) = \nu_{n}(-y_{0})$ rise sharply 
only in the vicinity of the edges $y_{0} \sim \pm W/2$, and 
recover the bulk values $n$ as $y_{0}$ moves to the interior 
a few magnetic lengths away from the edges;~\cite{MS} 
for the lowest $n=0$ level, $0\le \nu_{0} < 0.003$
for $y^{W}_{0} > 2.5\ell$.
Applying the Feynman-Hellman theorem to $\partial H_{0}/\partial
y_{0}$ reveals that a single electron state $N =(n,y_{0})$ 
has its center-of-mass at position
\begin{equation}
y^{\rm cm}_{n} \equiv \langle N|y|N \rangle 
= y_{0} - \ell^{2}\,  {\nu_{n}}'(y_{0})
\label{ycm}
\end{equation}
on the $y$ axis.
This $y^{\rm cm}_{n}$ stays within the interval $[-W/2,W/2]$ while 
$y_{0}=\ell^{2}\,p_{x}$ can take values outside it for electrons 
residing near the edges.

The normalized wave functions $\phi_{N}(y)$, taken to be real, are
functions of $y-y_{0}$ and $y^{W}_{0}$:
\begin{equation}
 \phi_{N}(y)= (1/\sqrt{\ell}) \phi_{n} \left( (y-y_{0})/ \ell;
y^{W}_{0}/ \ell\right),
\label{phiN} 
\end{equation}
where we have taken $\phi_{n}$ dimensionless so that its
dependence on dimensional quantities 
$\ell, y_{0}$, etc.~is made explicit.  
Let us denote the spatial distribution of a single electron state 
$N$ as
\begin{equation}
 \rho_{N}(y) \equiv \rho_{n}(y;y_{0}) ={1\over{\ell}}\,
\phi_{n} \left( (y-y_{0})/ \ell; y^{W}_{0}/ \ell\right)^{2}, 
\end{equation}
which is uniform in $x$ and which is normalized so that
\begin{equation}
 \int^{W/2}_{-W/2} dy\,\rho_{n}(y;y_{0})= 1.
\end{equation}
The associated current density per electron is written as a product of
charge $(-e)$, velocity $(p_{x}-eBy)/m$ and density $\rho_{n}/L_{x}$:
\begin{equation}
 j_{x}(y|N)={e\omega\over{L_{x}}}\, (y-y_{0})\, \rho_{n}(y;y_{0}).
\label{jxyN}
\end{equation}

The distributions $\rho_{n}(y;y_{0})$ are highly localized and are
nonzero only for $|y - y_{0}|\le O(\ell)$. 
In the sample bulk [i.e., $|y_{0}|< W/2 - O(\ell)$], in particular, 
$\rho_{n}$ are even functions of $y - y_{0}$:
\begin{equation}
\rho^{\rm bulk}_{n}(y - y_{0})= H_{n}(Y)^{2}\,
e^{-Y^{2}}/(\ell 2^{n}n!\sqrt{\pi}), 
\label{rhobulk}
\end{equation}
with $Y\equiv (y - y_{0})/\ell$.
The profile of $\rho_{n}(y;y_{0})$ gets gradually deformed 
as $y_{0}$ lies closer to the edge, 
as shown in Fig.~2 (a) for the $n=0$ level.
There the $y_{0}^{W}=3.5\ell$ electron state, depicted with a solid curve, has 
$\nu_{0}\approx 10^{-5}$ and thus exhibits essential characteristics 
of an electron in the bulk; clearly the associated current distribution,
depicted with a solid curve in Fig.~2 (b), 
is characteristic of an orbiting electron and 
carries no net current.
In contrast, for an electron lying closer to the edge 
the cyclotron motion gets deformed, yielding a net current,~\cite{H,MS}  
which, in view of Eq.~(\ref{ycm}),  is proportional to 
the gradient of the spectra $\nu_{n}$:
\begin{equation}
\int^{W/2}_{-W/2} dy\, j_{x}(y|N) 
= -{e \omega \over{L_{x}}} \  \ell^{2}\, {\nu_{n}}'(y_{0}) .
\label{intjx}
\end{equation}
The current-carrying property of each electron state $N$ is made
manifest by decomposing the current $j_{x}(y|N)$ 
into the circulating component $\propto (y - y^{\rm cm}_{n})\,
\rho_{n}(y;y_{0})$ (that carries no net current) and  
the traveling component $\propto
-\ell^{2} {\nu_{n}}'(y_{0})\, \rho_{n}(y;y_{0})$; 
see Fig.~2 (c) and (d).
It is clearly seen from these figures 
that the electrons residing near the edge (the edge states)  carry 
a large amount of current. Classically these edge states are 
visualized as electrons hopping along the sample edges.~\cite{Peierls}
They travel fast (with velocity $v_{x}=\omega \ell^{2}
{\nu_{n}}' \sim \pm \omega \ell $) in opposite 
directions at opposite sample edges.

Let us next examine the current distributions filled Landau levels 
support. We shall focus on the case of well-filled levels close to
integer filling.
It is convenient to use the Fermi potential 
$E_{\rm F}\equiv \omega (n_{\rm F} +1/2)$ or $n_{\rm F}$
to specify the filling of each level:
For a given $n_{\rm F}$ Landau levels $n$ with $n \le n_{\rm F}$ are  
populated densely with electron states over the range 
$(y_{0}^{-})_{n} < y_{0} < (y_{0}^{+})_{n}$ so that
$\nu_{n}((y_{0}^{\pm})_{n}) = n_{\rm F}$; 
the filling fraction of each level is 
$f_{n} = [(y_{0}^{+})_{n} - (y_{0}^{-})_{n}]/W$ with
the number of electrons given by $(N_{e})_{n}= (eB/2\pi)L_{x}W f_{n}$.
The $n_{\rm F}$ stays constant $n_{\rm F}=n$ while the $n$th level 
is gradually filled with the electron states residing in the sample 
bulk (the bulk states), and it increases as soon as the
edge states start to be filled. 
Such filled states $(n,y_{0})$ give rise to the current distribution
\begin{equation}
j_{x}(y)={e\omega\over{2\pi
\ell^{2}}} \sum_{n} \int_{(y^{-}_{0})_{n}}^{(y^{+}_{0})_{n}}
 d y_{0}\, 
(y - y_{0})\, \rho_{n}(y;y_{0}).
\end{equation}
In view of Eq.~(\ref{rhobulk}), it is clear that $j_{x}(y)=0$ 
in the sample bulk.
Figure~3(a) shows the current distribution near the lower edge 
$y=-W/2$ for various values of $n_{\rm F}$.
For $0<n_{\rm F}<1$ only the $n=0$ level carries current,
for $1<n_{\rm F}<2$ two levels ($n=0,1)$ contribute, and so on.

There are two notable features read from the figure: 
(1) The current flows along the edge in a channel
whose width $\sim O(\ell)$ increases with the number of
current-carrying levels.  
This feature has been well known~\cite{H,MS} and is naturally expected
from the current fraction per electron in Eq.~(\ref{intjx}).
Somewhat unexpected is the following: 
(2) Both the spatial distribution and direction of the current change
dramatically with $n_{\rm F}$. 
For small $n_{\rm F}\sim 0.1$ the edge current flows predominantly 
in the negative $x$ direction while a large fraction of current flows
in the opposite direction for $n_{\rm F}\sim 0.9$; the
integrated amount of current changes its sign 
at $n_{\rm F}=0.5$.  Analogous patterns of current flow are also seen 
for each integer interval $n < n_{\rm F} < n+1$.
The amount of current integrated over one edge oscillates with
$n_{\rm F}$, as shown in Fig.~3(b).

These changing and oscillating current distributions may appear puzzling
if one notes Eq.~(\ref{intjx}) alone.
There is, however, a simple resolution~\cite{Peierls}: 
Associated with each orbiting electron is a circulating current 
$\propto (y - y^{\rm cm}_{n})\, \rho_{n}(y;y_{0})$ that
produces magnetic moment pointing opposite to the applied magnetic
field $B$.  For a dense collection of electrons in the sample interior 
this circulating 
diamagnetic current is averaged out to vanish locally, but leaves 
a current circulating along its periphery.
Indeed, collecting the circulating current components alone
yields a current distribution $j_{x}^{\rm circ}(y)$ localized 
near the edge, as shown in Fig.~4 (a) for the $n=0$ level with 
different $n_{\rm F}$.
Also shown in Fig.~4 (b) is the current distribution 
$j_{x}^{\rm tr}(y)$ formed by the traveling current components 
alone for the same $n=0$ level.
They combine to build up the current distributions of Fig.~3(a).
It is now clear that the current distribution for $n_{\rm F}= 0.1$ 
in Fig.~3 (a) primarily derives from the orbital diamagnetic current 
whereas the edge-driven traveling current dominates 
in the $n_{\rm F}= 0.9$ case. 
Note that the edge-driven current works to cancel 
the orbital diamagnetic current. 
The cancellation is partial in quantum theory,~\cite{Peierls} 
leading to the Landau diamagnetism of electrons 
for three-dimensional samples.

In the present two-dimensional case it is possible to calculate 
both the diamagnetic and  edge-driven components of the edge current
explicitly. As a preliminary step, let us first consider 
a collection of 
electron bulk states (of the $n$th level) that
fill up a half-infinite 
interval $y_{0}\ge 0 $. They lead to the current distribution
\begin{equation}
j_{x}^{\rm bulk}(y) = - {e\omega\over{2\pi \ell^{2}}} 
\int^{\infty}_{-y}dz\, z \rho^{\rm bulk}_{n}(z)
 \label{bulkcurrent}
\end{equation} 
localized around $y= 0$ with spread $\triangle y = O(\ell)$; 
for $n=0$ this is given by $-(2\sqrt{\pi})^{-1}
e^{-y^{2}/\ell^{2}}$ times $(e\omega/2\pi \ell)$.
The integrated amount of this diamagnetic current increases with $n$:
\begin{equation}
\int_{y\sim 0} dy\,j_{x}^{\rm bulk}(y) = 
- \left(n+{1\over2}\right) {e\omega\over{2\pi}}.
  \label{jdiamag}
\end{equation}
Consider next a collection of electron states that fill up the interval 
$(y_{0}^{-})_{n} \le y_{0} \le \eta$ with $\eta$ lying somewhere 
far in the sample interior. 
The circulating current components associated with
these states, though carrying no net current, combine to build up 
two prominent current distributions localized near the edge $y= -W/2$
and around $y=\eta$, which are equal in net amount of current and
opposite in sign.  
This shows that the diamagnetic component of the current 
localized near the edge $y= -W/2$ supports a fixed amount~\cite{fn} 
equal to $ - (e\omega/2\pi) (n + 1/2)$, irrespective 
of its distribution as well as the shape of edge potentials.
In particular, the four different distributions of 
$j^{\rm circ}_{x}(y)$ in Fig.~4(a) support the same amount
of current equal to $-(e\omega/4\pi)$.

On the other hand, the edge states of the $n$th level arise for
$n_{\rm F} > n$ and, as seen from Eq.~(\ref{intjx}), carry the amount 
of current equal to $(e\omega/2\pi)(n_{\rm F} - n)$ at the lower edge
$y\sim -W/2$. With the two current components put together 
the $n$th level alone supports the amount of edge current
\begin{equation}
J^{\rm edge}_{n} = {e\omega\over{2\pi}} 
\left(n_{\rm F}- 2n -{1/2}\right)\,
\ \ \ \ \ \ \ \ \ \mbox{for $n_{\rm F}\ge n$}.
\label{Jnedge}
\end{equation}
For the integer interval $n\le n_{\rm F} < n+1$ the lower
$(n+1)$ filled levels combine to support the amount of
current 
\begin{equation}
 J^{\rm edge}={e\omega\over{2\pi}}
(n+1)\left[ n_{\rm F}- (n +{1\over2})\right],
\label{Jedge}
\end{equation}
localized near the edge $y = -W/2$.
This current-$n_{\rm F}$ relation accounts for the alternating edge
current of Fig.~3(b), and shows that the (integrated) edge current
changes its direction precisely at $n_{\rm F}= n + 1/2$.

The alternating edge current of Fig.~3 is intimately connected to an
oscillatory variation of magnetic susceptibility of an electron gas
with varying magnetic field, known as the de~Haas -- van~Alphen 
(dHvA) effect.~\cite{Peierls,Ishihara}
Indeed, the current $J^{\rm edge}$, circulating along the sample edge, 
gives rise to uniform magnetization normal to the sample plane and of
magnitude (per unit area)
\begin{equation}
M_{z}=\mu_{\rm B} (1/2\pi \ell^{2}) (n+1)[2(n_{\rm F} -n) - 1]
\label{Mz}
\end{equation}
for $n\le n_{\rm F} < n+1$, apart from corrections of $O(\ell/W)$ 
that vanish as $W\rightarrow \infty$, where $\mu_{\rm B}= e \omega
\ell^{2}/2 = e/(2m)$ is the Bohr magneton (with effective mass $m$).

This result is readily verified by thermodynamic methods as well. 
The simplest way is to consider 
how the energy of the present finite-width system,
\begin{equation}
E= (L_{x}/2\pi \ell^{2}) \sum_{n}\int d y_{0}\,
\omega\,\left\{ \nu_{n}[y^{W}_{0}/\ell] + 1/2\right\},
\end{equation} 
responds to an infinitesimal variation of the magnetic field $B$.
Keep $ p_{x}= y_{0}/\ell^{2}$ fixed and calculate the magnetization 
$M_{z}= - (1/WL_{x})\delta E/\delta B$:
\begin{equation}
M_{z} = -{\mu_{\rm B} \over{2\pi \ell^{2}}}\sum_{n} \int {d 
y_{0}\over{W}} \,
\left\{ 2\nu_{n}(y_{0}) + 1  -  
(2y_{0} - y^{W}_{0}) \nu_{n}'(y_{0}) \right\}.
\label{Mzyzero}
\end{equation}
Note that $\nu_{n}=n$ in the sample bulk and that $\nu_{n}'\not=0$ 
only in the edge regions. 
The $y_{0}$ integral therefore is equal to $[2n+1 -2(n_{\rm F} -n)]$ 
for each filled level $n$, apart from corrections of $O(\ell/W)$; 
this result is consistent with Eq.~(\ref{Jnedge}) and 
thus recovers Eq.~(\ref{Mz}).
It is also enlightening to confirm Eq.~(\ref{Mzyzero}) by a direct
calculation of the thermodynamic potential $\Omega(\mu,T;B)= -T
\sum_{i} \ln(1 + e^{(\mu - \epsilon_{i})/kT})$ with $T\rightarrow 0$.

With the spectra $\nu_{n}(y_{0})$ determined numerically, 
it is a simple task to express the magnetization (\ref{Mzyzero}) 
as a function of $B$ or the filling factor 
$f = \sum_{n} f_{n} \propto N_{e}/B $.
As seen from Fig.~5, 
the magnetization per electron, $m_{z}\equiv M_{z}/(N_{e}/WL_{x})$,
plotted as a function of $f$ 
shows an oscillatory variation characteristic of the dHvA effect. 
As usual, the gradual variation of $m_{z}$ in each integer interval 
$n < f <n+1- O(\ell/W)$ is ascribed to the diamagnetism of 
orbiting electrons in the bulk.
In each tiny interval $n+1 - O(\ell/W) < f <n+1$,  $m_{z}$ makes 
a drastic change from diamagnetism to paramagnetism that is caused 
by the edge states of the $n$th level.

Note here that the electron edge states, because of their tiny filling
fraction$\sim O(\ell/W)$, scarcely contribute to the internal 
energy $E$.
Thus, even without explicit account of the edge states, 
one can still arrive at the dHvA effect by thermodynamic methods
(i.e., through $E$ or $\Omega$), except that $M_{z}$ now shows
discontinuous jumps at integer filling $f=n$. 
The edge states are invisible in such derivations but are certainly
present physically: 
Note that magnetization by nature is continuous as a
function of $B/N_{e}$, as seen clearly from the $W=30\ell$ case of
Fig.~5.  
In view of this continuity, 
the prominent sign change of $m_{z}$ near integer filling $f=n$
(which can be derived by thermodynamics without the edge 
states) does imply the presence of the edge states and, in particular,
the alternation in direction of the edge current.
In this sense, the alternating edge current in Fig.~3 is
another aspect of the dHvA effect.

The alternation of the edge current is caused by competition between
the circulating and edge-driven components of the current 
near the edge.  The edge-driven component is naturally 
very sensitive to the shape of the edge potential while the 
circulating component is not; see Fig.~2 (c) and (d).  
Correspondingly the edge current in general varies in
distribution according to the shape of edge potentials.
Still, in net amount per edge, both current components 
remain unaltered
so that Eqs.~(\ref{Jnedge}) and (\ref{Jedge}) hold,
irrespective of the details of edge potentials.  \\

\setcounter{equation}{0}
\setcounter{section}{3}
\noindent
{\bf 3.\ \  Hall-current distributions}
\medskip

The alternation of the edge current, which we have just seen, 
does not necessarily imply that the information the edge current
carries also changes the direction of propagation.  
In this section we clarify this point by studying 
how the edge current responds to a Hall field.

For simplicity let us consider a uniform field 
$E_{y}= -\partial_{y}A_{0}(y)$.
Its effect is readily taken care of by making the shift
\begin{equation}
y_{0} \rightarrow 
\bar{y}_{0} \equiv y_{0} - (e\ell^{2}/\omega) E_{y}
\end{equation}
in $H_{0}$ of Eq.~(\ref{hzero}),
and the full Hamiltonian $H_{0} -eA_{0}(y)$ leads to the 
new spectrum $\epsilon_{n}(\bar{y}_{0}) + e E_{y}\bar{y}_{0}
+O(E_{y}^2)$.
The normalized wave function $\phi_{N}(y)$ 
is given by Eq.~(\ref{phiN}) 
with the replacement $y_{0}  \rightarrow  \bar{y}_{0}$, and
the current per electron $j_{x}(y|N)$ by Eq.~(\ref{jxyN}) with 
$\rho_{n}(y;y_{0})  \rightarrow \rho_{n}(y;\bar{y}_{0})$.
The current distribution $j_{x}(y|N)$ itself barely changes thereby
because, under conditions of practical interest, 
the deviation $\bar{y}_{0}-y_{0}$ is negligibly small.

Actually the Hall current we are interested in is the small deviation,
\begin{equation}
\triangle j_{x}(y|N)={e\omega\over{L_{x}}}\, (y-y_{0})\, \left\{
 \rho_{n}(y;\bar{y}_{0})- \rho_{n}(y;y_{0}) \right\},
\label{trijxyN}
\end{equation}
representing the response to an applied field.
As seen from Fig.~6(a), unlike $j_{x}(y|N)$, 
the Hall current per electron
$\triangle j_{x}(y|N)$ is primarily composed of traveling components.
Here we see that, while the edge states and bulk states are drastically
different in the amount of current they carry, they are essentially 
the same in the Hall-current transport.
Numerically an edge state supports even a smaller amount of
Hall current than a bulk state, as seen from Fig.~6(b), where
the numerically-integrated amount of Hall current 
per electron, $\triangle J_{x}\equiv \int dy\, \triangle
j_{x}(y|N)$,  is compared with the net current per electron,  
$J_{x}\equiv \int dy\, j_{x}(y|N) \propto -{\nu_{n}}'(y_{0})$.

It is possible to understand the current-carrying properties of 
each electron state in a more general way:
Of the current $j_{x}(y|N)$, the circulating component 
$\propto (y- y^{\rm cm}_{n})\, \rho_{n}$ carries no net current.
It is the traveling component, associated with the drift of 
a Hall electron with velocity
\begin{equation}
\omega \ell^{2}\, {\nu_{n}}'(\bar{y}_{0}) + E_{y}/B,
\label{vdrift}
\end{equation}
that carries a net current (and hence information with it).
Accordingly disturbances caused upon a sample,
 e.g., by varying a magnetic field 
or electron population will propagate in a direction fixed by the
edge with velocity$\sim \omega \ell$.
In contrast, the effect of a Hall field (i.e., the Hall current)
propagates in a direction fixed by
the polarity of $E_{y}$ with velocity $\sim E_{y}/B$.
[Numerically $\omega \ell \sim 10^{7}$~cm/s for
typical values $\omega \sim 10~$meV and $\ell \sim$ 100~{\AA} while 
$E_{y}/B \sim 10^{3}$~cm/s for $E_{y}$ = 1~V/cm and $B$ = 5~T.] 
It follows from Eq.~(\ref{vdrift}) that the net amount of 
Hall current per electron is proportional to 
$1 - \ell^{2}\, {\nu_{0}}''(y_{0})$, which reproduces 
the numerical result in Fig.~6(b) very accurately.

Filled Landau levels support the prominent edge-current distributions 
of Fig.~3(a), which remain essentially unchanged in the presence of 
a Hall field as well.
These edge currents flow in opposite directions at the two opposite
edges of a sample, and in equilibrium with $E_{y}=0$ 
they combine to vanish.~\cite{H,MS} 
When a Hall field $E_{y}$ is turned on, the Hall current emerges as a
small deviation from the equilibrium distribution,
as shown in  Fig.~6(c) for the present impurity-free and 
uniform-$E_{y}$ case. 
It is clear from Fig.~6, contrary to some tempting idea~\cite{B}, 
that the electron edge states, because of their tiny filling 
fraction $\sim O(\ell/W)$, share only a tiny portion 
of the Hall current.~\cite{KS}

\setcounter{section}{4}
\noindent
{\bf 4.\ \  Concluding remarks}
\medskip

In this paper we have examined current distributions in a Hall bar
in the regime of the integer QHE, and shown, 
in particular, that the edge current changes its distribution 
and direction as the number of electron edge states increases.
This dramatic change is a consequence of competition between the
circulating diamagnetic component and edge-driven traveling
component of the current carried by electrons near the sample 
edge, and is closely related to the dHvA effect.
It should be emphasized that the dHvA oscillation of magnetization is 
indirect evidence for the edge states; it is clear physically that 
without them no oscillation would arise, since orbital magnetization
alone leads to diamagnetism.~\cite{Peierls} 
In this connection we have seen explicitly that
magnetization is continuous as a function of $B/N_{e}$ when the edge
states are properly taken into account.
We have also seen that the Hall current flows 
in a manner quite independent of the equilibrium edge-current 
distribution.

Finally we would like to comment on the influence of disorder.
Rapid motions of Hall electrons like cyclotron motion and 
acceleration by the edge are potentially not very sensitive to
disorder. 
The edge current of Fig.~3, resulting from such rapid motions,
will therefore remain prominent in the presence of weak disorder as
well, and continue to alternate in direction 
as the number of edge states increases. 
In contrast, disorder will substantially modify the current carried 
by slowly-drifting electrons in the sample bulk: 
In the presence of disorder a large fraction of electron bulk states 
get localized and cease to carry current; at the same time, 
the surviving extended states carry more current 
and achieve exact compensation.~\cite{AA,Pr}  
This exact current compensation is a consequence of electromagnetic
gauge invariance and takes place under general circumstances
involving both bulk and edge states.~\cite{KS}  

Note now that, of the electron bulk states, 
those residing near the edge of 
the sample bulk (``bulk edge''), though influenced by disorder, would 
have a better chance of staying extended than those far in the bulk.
In view of current compensation, a considerable portion of the Hall
current, redistributed via disorder, would therefore flow along 
the sample edges. 
In this way it has been pointed out~\cite{KS} that there are two kinds, 
fast and slow, of edge current in Hall samples in the regime of 
the integer QHE.

These two kinds of edge current differ in channel width and 
in direction of flow. The fast component is nothing but the 
alternating edge current discussed so far, consisting of the two
(circulating and traveling) subcomponents; 
it has a channel width of $O(\ell)$ and flows in opposite directions
at opposite sample edges.
In contrast, the slow component, the ``bulk-edge'' Hall current, 
will have a channel width, related to some localization length
characteristic to the bulk edge, 
which could well be larger than $O(\ell)$. 
This slow edge current will flow in the same direction at opposite
edges and reverse direction when the polarity of the Hall field is
flipped; an observed Hall-potential distribution~\cite{FKHBW} appears 
to be in support of this feature.

A  numerical experiment is now under way to study the current
distribution for small samples with randomly distributed impurities,
and is yielding results that appear to confirm the effect of disorder 
on the current distribution described above; details will be reported
elsewhere.\\

{\bf Acknowledgments}

The author wishes to thank B. Sakita and Y. Nagaoka for
useful discussions. 
This work is supported in part by a Grant-in-Aid for 
Scientific Research from the Ministry of Education of Japan, 
Science and Culture (No. 07640398).   \\

%%%%%%%%%%%%%%%%%%%%% Figure captions %%%%%%%%%%%%%%%%%%%%%%%
\newpage

\begin{figure}
\vskip2cm
\epsfxsize=14cm
\epsfysize=9cm
\centerline{\epsfbox{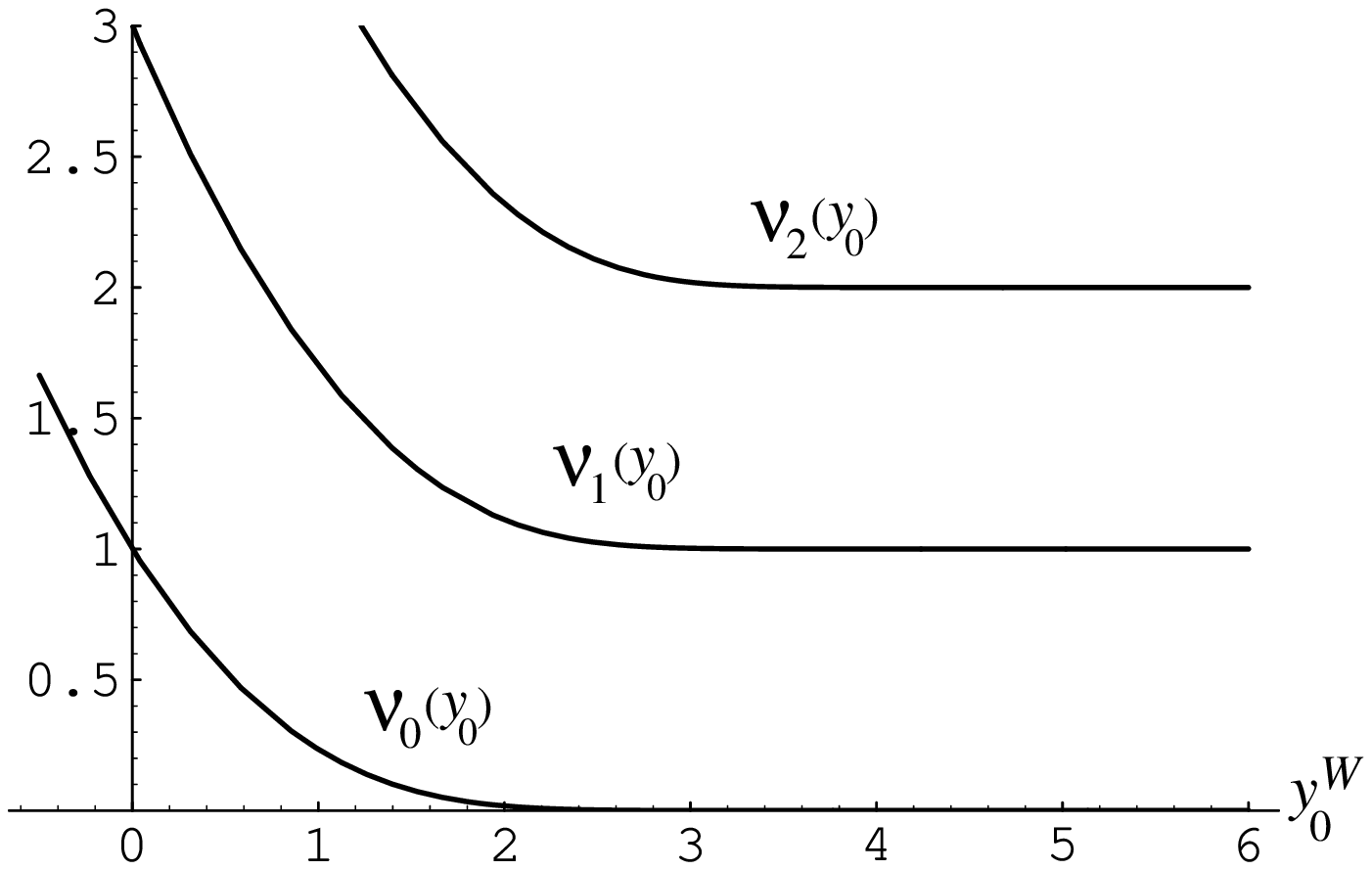}}
\vspace{3cm}
\caption{Spectra $\nu_{n}(y_{0})$ plotted as a function of 
$y_{0}^{W}\equiv  y_{0} + W/2$ (measured in units of the magnetic
length $\ell$) near the lower edge $y_{0}^{W}\approx 0$.}
\label{fig1}
\end{figure}

\newpage
\vskip2cm~

\begin{figure}
\epsfxsize=11cm
\epsfysize=14.5cm
\centerline{\epsfbox{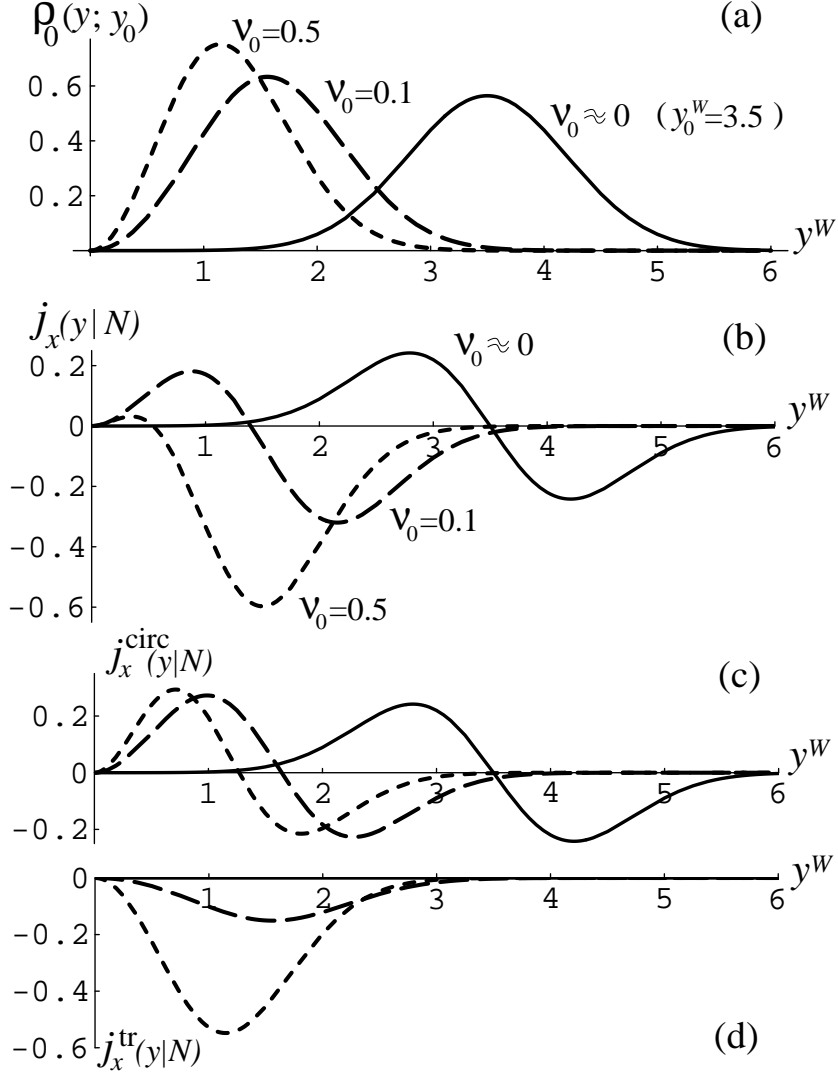}}
\vspace{2cm}
\caption{ (a) Spatial distributions  $\rho_{n}(y;y_{0})$ of 
single electron states with different $\nu_{0}(y_{0})$ 
in the $n=0$ level.  
The solid curve represents
the $\nu_{0}\approx 10^{-5}$ electron state with $y_{0}^{W}=3.5 \ell$, 
the long-dashed curve the $\nu_{0}=0.1$ state with 
$y_{0}^{W}\approx 1.40 \ell$, and the dashed curve 
the $\nu_{0}=0.5$ state with $y_{0}^{W}\approx 0.541 \ell$.
The coordinate $y^{W}= y + W/2$, defined relative to the edge 
$y= -W/2$, is measured in units of $\ell$.
(b) Distributions of the associated current per electron.  
Here, for convenience, the current $j_{x}(y|N)$ is plotted
in units of $-e \omega/L_{x}$ so that its sign refers to that 
of the velocity $v_{x}$ along the $x$ axis.
(c) Decomposition of $j_{x}(y|N)$ into the circulating component 
$j_{x}^{\rm circ}(y|N)$ and 
the traveling component $j_{x}^{\rm tr}(y|N)$.}
\label{fig2}
\end{figure}

\newpage
\vskip2cm~

\begin{figure}
\epsfxsize=10.5cm
\epsfysize=16cm
\centerline{\epsfbox{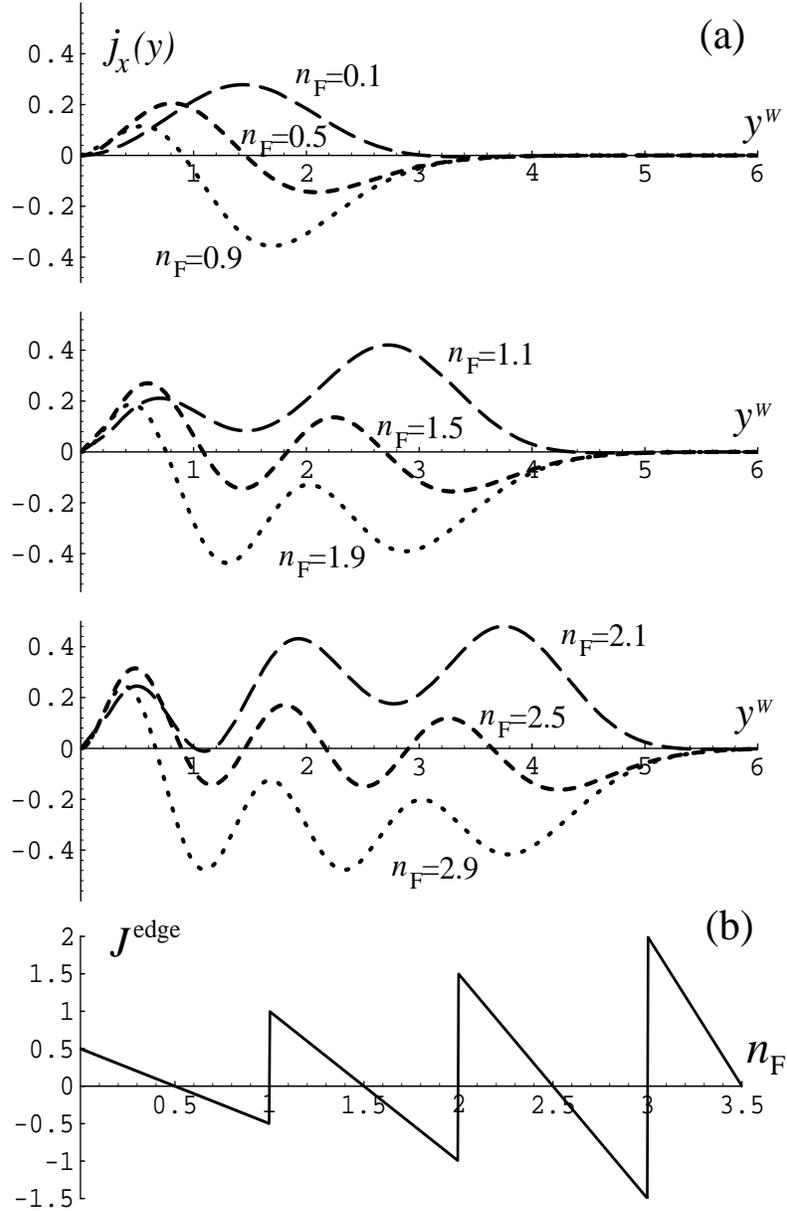}}
\vspace{3cm}
\caption{
(a) Current distributions filled Landau levels support near the sample 
edge $y = - W/2$;  $n_{\rm F}=E_{\rm F}/\omega -1/2$ refers to the 
Fermi potential $E_{\rm F}$.
The current $j_{x}(y)$ is measured in units of $-e\omega/(2\pi)$ and
the coordinate $y^{W}= y + W/2$ in units of $\ell$.
(b) The integrated amount of the current per edge [in units
of  $-e\omega/(2\pi)$] oscillates with $n_{\rm F}$.}  
\label{fig3}
\end{figure}

\newpage
\vskip2cm~

\begin{figure}
\epsfxsize=12.5cm
\epsfysize=10cm
\centerline{\epsfbox{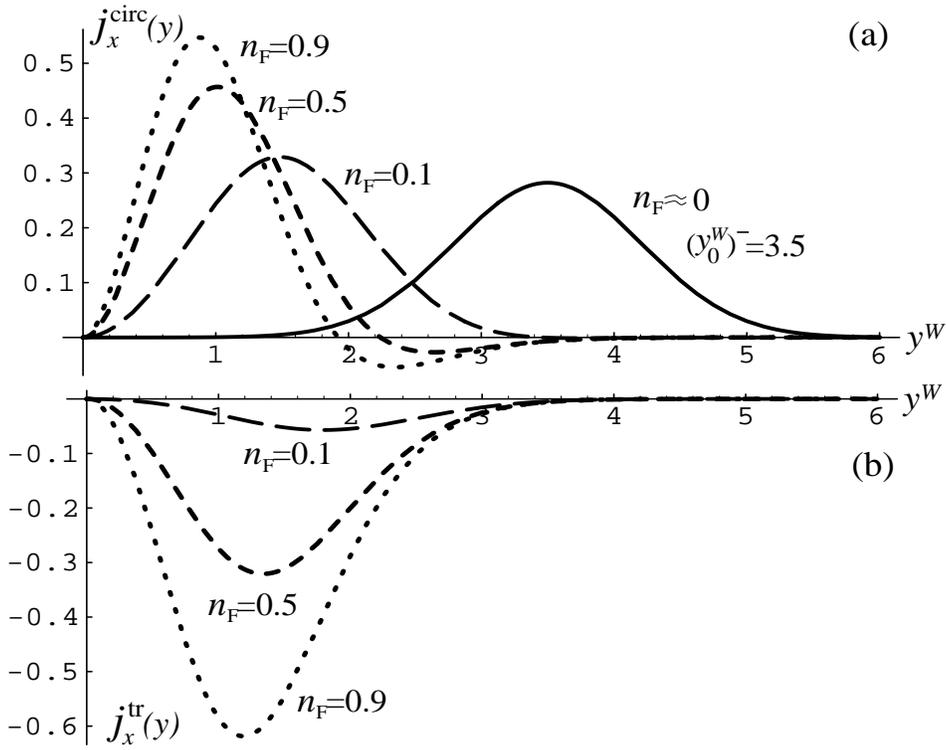}}
\vspace{3cm}
\caption{(a) Current distributions near the edge 
$y = - W/2$ built up by the circulating current components alone 
for the $n=0$ level with different values of $n_{\rm F}$.
The $n_{\rm F}\approx 10^{-5}$ case refers to the $n=0$ level 
filled with electron states $y_{0}^{W}\ge 3.5 \ell$.
The current is measured in units of $-e\omega/(2\pi)$, and
the coordinate $y^{W}$ in units of $\ell$.
(b)  Current distributions formed by the traveling current
components alone for the same $n=0$ level.}  
\label{fig4}
\end{figure}

\newpage

\vskip3cm~

\begin{figure}
\epsfxsize=10cm
\epsfysize=6cm
\centerline{\epsfbox{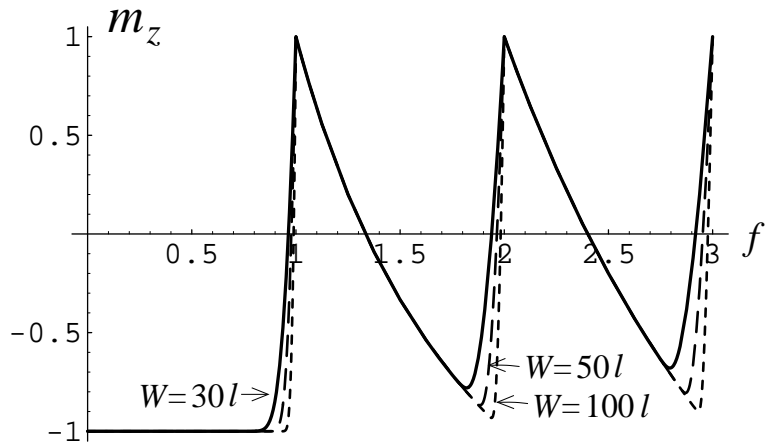}}
\vspace{3cm}
\caption{ Magnetization per electron 
[in units of $\mu_{\rm B}/2 \pi \ell^{2}$]
v.s.~the filling factor $f$ for a Hall bar of
finite width $W = 30 \ell, 50 \ell$, and $100 \ell$.}
\label{fig5}
\end{figure}

\newpage
\vskip2cm~

\begin{figure}
\epsfxsize=10.5cm
\epsfysize=14.5cm
\centerline{\epsfbox{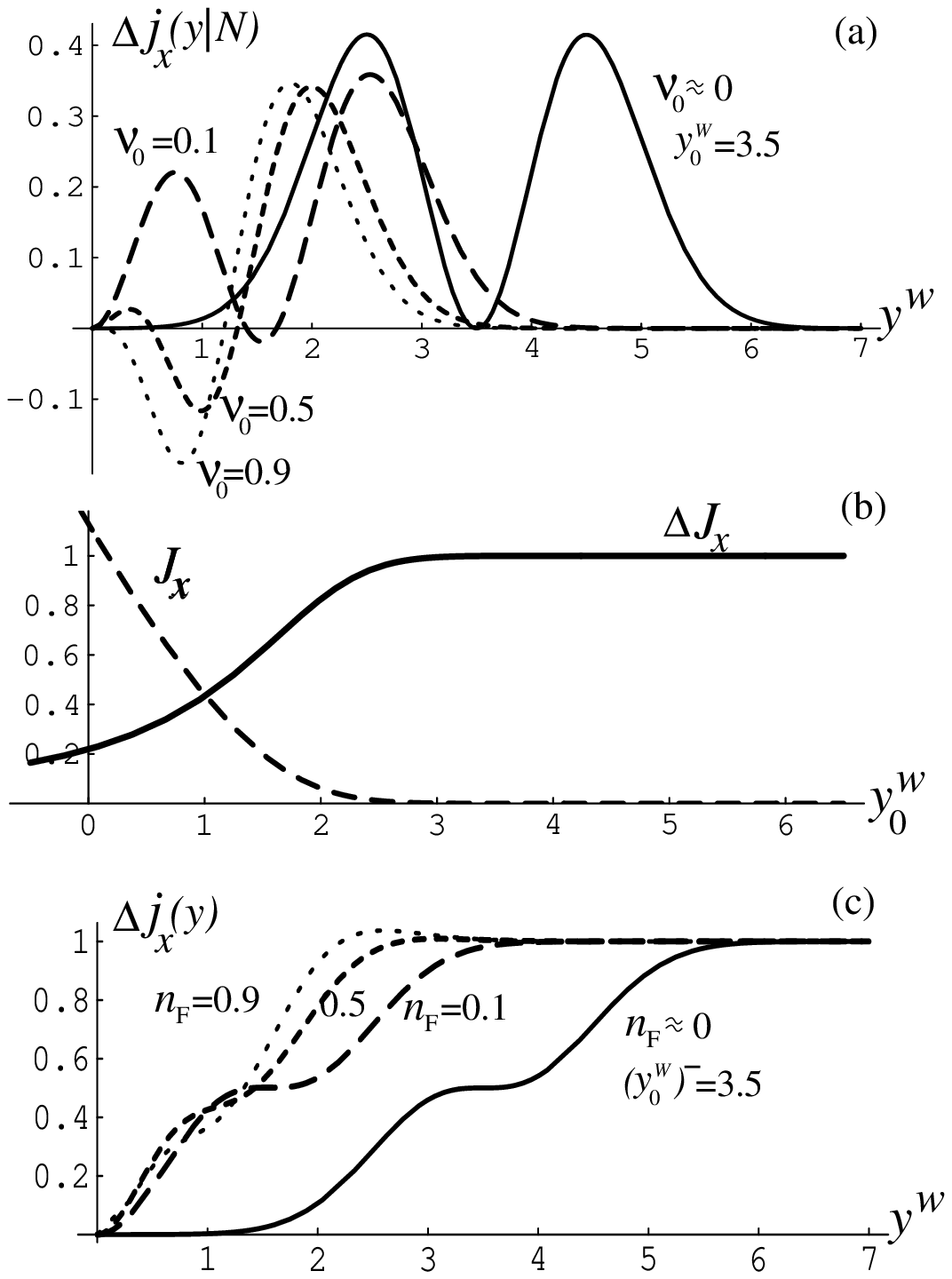}}
\vspace{3cm}
\caption{ (a) Distributions of the Hall current per electron, 
$\triangle j_{x}(y|N)$, for electron states with different $\nu_{0}(y_{0})$ 
in the $n=0$ level.  The current is plotted in units 
of $ 10^{-3}\times (-e \omega /L_{x})$
for $e\ell E_{y}/\omega = 10^{-3}$.  
(b) The Hall current per electron 
$\triangle J_{x}= \int dy\, \triangle j_{x}(y|N)$ 
(in units of $ - e^{2}\ell^{2}E_{y}/L_{x}$) for
electron states in the $n=0$ level, 
in comparoson with the net current per electron 
$J_{x}= \int dy\, j_{x}(y|N)$ (in units of $e \omega \ell/L_{x}$).
(c) Hall-current distributions near the edge for the $n=0$ level.
The edge states increase in number with $n_{\rm F}$.}
\label{fig6}
\end{figure}


\begin{thebibliography}{99}
\bibitem{fn} This value $(e\omega/2\pi) (n + 1/2)$ for the edge
current was derived earlier in Ref.~19  within a
derivative expansion for a slowly-varying confining potential.
\end{thebibliography}
\end{document}